\documentclass[aps,prl,twocolumn,floats,showpacs]{revtex4}
\usepackage{graphicx}
\begin{document}
\title{Impact of Coulomb interaction and Kondo effect on transport in quantum dots}
\author{A. Golub}

\affiliation{ Department of Physics, Ben-Gurion University of the
Negev, Beer-Sheva, Israel \\   }
\begin{abstract}
We examine the impact of Coulomb electron-electron interaction on
transport in a junction with a quantum dot described by Kondo
Hamiltonian. We analyze the Fermi liquid regime and consider the
limit of zero temperature. With the help of Keldysh technique we
calculate the non-linear current and shot noise as a function of
applied voltage. We show that Coulomb interaction markable
influences the universal effective charge of current-carrying
particles $e^*=5/3e$ which can be measured in shot-noise
experiments. The
 electron-electron interaction modifies this universal value by a factor
 ($e^*=5/3eF$) which is  less then one and also voltage dependent.
\end{abstract}
\maketitle

 The non-linear conductance and
zero-frequency noise out of equilibrium (shot noise) can yield
information on charge fluctuations in physical systems
\cite{oreg}. Indeed, in many cases the effective charge of
carriers may be different from bare electron charge $e$. This
occurs when the fundamental quasiparticle charge does not coincide
with $e$ or due to interactions. The Kondo effect in the quantum
dots opens the opportunity to investigate the changing of
effective charge due to interactions. For this purpose it is of a
special interest the Fermi liquid regime of Kondo model. In this
case the backscattering current is defined by the scattering term
and by the interaction quadratic in the spin current. The
measurements of the backscattering current and shot noise can give
the information about effective backscattering charge of the
carriers influenced by interactions. These reasons have been put
forward in a recent work \cite{oreg} where the effective
current-carrying charge of the particles was found to acquire the
universal value $e^*/e =5/3$.  However, as we will show in the
present article, the Coulomb interaction and corresponding
electrical potential fluctuations have  a noticeable impact on
this effective charge. They change the value of backscattering
charge and cause a voltage dependence of $e^*$. To obtain the
effective charge with or without electrical potential fluctuations
we have to find  both parts (scattering and interaction) of the
current and shot noise \cite{oreg,gogolin1,me}. We concentrate in
the following on $T=0$ limit, where the thermal noise vanishes. We
consider the regime around the strongly interacting fix point
(Fermi liquid regime) of Kondo Hamiltonian, that is, $eV<< T_K$
($T_K$ is the Kondo temperature) counting for potential
fluctuation at the one loop approximation. Here we notice that the
role of Coulomb interaction in the case of a large dots was
analyzed before in a number of articles
\cite{gol,aleiner,brouwer}.

Our principal result can be written compactly as
$e^*/e=\frac{5}{3}F$, where factor $F$ depends on the bias
voltage, charging energy $E_C =e^2/C$ (here $C$ is the total
capacitance), and the effective number of conducting channels
$\bar{N}$. In Fig.1 we plot this factor as a function of voltage
for different $\bar{N}$. For validity of perturbation theory that
we use, the absolute value of the voltage must be not too small.
Also for dot systems $E_C>> T_K$. In this limit we have
\begin{figure}
\begin{center}
\includegraphics [width=0.4 \textwidth ]{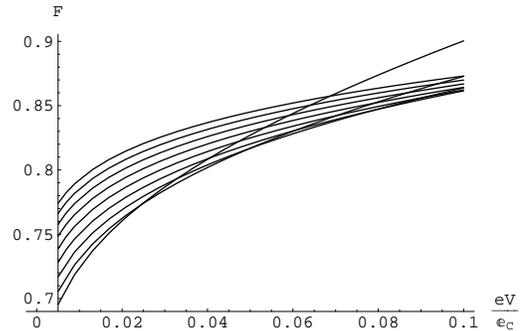}
\caption {Voltage dependence (in units of Coulomb energy $E_C$) of
factor F. The curves are related to different number of effective
conducting channels $\bar{N}$ starting from $\bar{N}$=2 up to
$\bar{N}$=10 }
\end{center}
\end{figure}
\begin{eqnarray}
F&=&\frac{1+a_n}{1+a_j}\label{factor}\\
a_n&=&\frac{4}{5\bar{N}}(\ln(\frac{2E_C \bar{N}}{\pi eV}+0.36)\\
a_j&=&\frac{7}{4\bar{N}}(\ln(\frac{2E_C \bar{N}}{\pi eV}+0.16)
\end{eqnarray}
In the region of small voltages factor $F$ (Fig1.) rather strongly
(~30\%) reduces  the effective charge. Now we present the basic
steps which lead to the derivation of equation (\ref{factor}).

  {\it Effective action}: Coulomb
interaction in a quantum dot
 is responsible for formation of Kondo effect
and for  electric potential fluctuations in the whole junction.
The Kondo effect depends also on the tunnelling rates between the
dot and the leads. The later strongly influence the Kondo
temperature $T_K$. The Coulomb interaction is characterized by the
charging energy $E_C $. The parameter that controls the impact of
electron electron interactions on the tunnelling is the total
conductance $g$ or the number of  open channels $N$. More
precisely, in the case of large $N$ we can find the interaction
corrections by performing $1/N$ expansion. Here we address the
unitary limit of Kondo quantum dot with odd numbers of electrons
in it, that is, we deal with ordinary Kondo effect in a quantum
dot. Though, in this case $N=2$ (effective $N$ may be increased by
a shunting resistance) we believe that even at $1/N$ level the
qualitative picture is correct. The problem is also interesting
because of its relation to a rare case of a non-equilibrium system
with interaction where the electrical potential fluctuations are
included. As an advantage of the unitary limit  is that the
effective transmissions are close to the unity. Indeed,
conductance in additions to $\sigma_0 = 2e^2/h$ has a small
nonlinear contribution $\sigma_V \sim -V^2/T_K^{2}$, where the
bias voltage $V<<T_K $. As we will show the Coulomb interaction
modifies only the nonlinear part of the transport current and shot
noise. Therefore, to the lowest order in $V/T_K$ one can calculate
the effective action for electric potentials taking in
considerations only the unitary limit. In this limit we can use
two approaches which give the same result for spin $1/2$ Kondo
model: one is based on scattering representation of the Kondo
Hamiltonian \cite{glazman}, the other is the mean field slave
boson approximation (SB) \cite{hew}. The SB is more transparent
and it will be used to obtain the effective action. We start with
the Anderson Hamiltonian
\begin{eqnarray}
\hat{H}&=&H_{L}+H_{R} + \!\sum_{k,\sigma,\alpha}(v_\alpha
c^{\dagger}_{\alpha \sigma,k}d^{\phantom{\dagger}}_{\sigma}+ {\rm
H.c.}
)\nonumber\\
&+&\sum_\sigma \epsilon d^{\dagger}_{\sigma}
d^{\phantom{\dagger}}_{\sigma}+
Ud^{\dagger}_{\uparrow}d^{\phantom{\dagger}}_{\uparrow}
d^{\dagger}_{\downarrow}d^{\phantom{\dagger}}_{\downarrow}\
\label{HA}
\end{eqnarray}
The first two terms correspond to non-interacting electrons in the
two leads
\begin{equation}\label{lead}
    H_{L(R)}=\sum_{k,\sigma}\xi_{L(R)k}c^{\dagger}_{L(R)\sigma,k}c_{L(R)\sigma,k}
\end{equation}
 where $c_{\alpha\sigma,k}$, $ \xi_{\alpha k}$
are the electron field operator  and the electron energy of a
lead. Index $\alpha=L,R$ indicates left (right) lead. We assume
that the leads are biased by the fluctuating electric field
potential drop $v(t)=\frac{d\varphi(t)}{dt}$. There are also
electron electron interactions in the leads which are not included
in the Hubbard coupling $U$. At first we separate the Hubbard
interaction that leads to Kondo effect. For this we apply to
Hamiltonian (\ref{HA}) the SB approximation at
$U\rightarrow\infty$. In the slave-boson approach, the localized
electron operator $d^{\dagger}_{\sigma}$ is represented by
$\hat{f}^\dagger_{\sigma}\hat{b}$ with $\hat{b}$ and
$\hat{f}^\dagger_{\sigma}$ being, respectively, the standard boson
and fermion operators. The action on Keldysh contour then will
incorporate the Anderson Hamiltonian (\ref{HA}) with the new boson
and fermion operators. The quadric term is replaced by  the two
new terms: one represents the renormalized level position in the
dot $\tilde{\epsilon}$, and the second stands for requirement of a
single occupancy which demands a  Lagrange multiplier $\lambda$
\begin{equation}\label{so}
  \hat{b}^\dagger\hat{b}+
  \sum_{\sigma}\hat{f}^\dagger_{\sigma}\hat{f}_{\sigma}=1
\end{equation}
  In the SB approximation  the Bose
 operators $\hat{b}^\dagger, \hat{b}$  are replaced by their expectation value
 $b$. Having the Anderson part of the action arranged we can
 get the general form of the total effective action by including the
 Coulomb  interaction term. Performing
 the standard Hubbard-Stratonovich decoupling of this term in the action, one
 introduces on the Keldysh contour fluctuating electrical
 potentials $v_i(t)=\frac{d\varphi_i (t)}{dt}, (i=1,2)$
 \cite{gol,kamenev}.  Next  we integrate out the fermion operators
  in the leads, and also add the action of external circuit expressed in terms of shunt
  resistance $R_{s}$. Thus the effective action acquires a form
    $S_{eff} =S_b+S+S_{s}$,
 where nonoperator bosonic part of the action is
\begin{eqnarray}
S_b=-\int\lambda_i(b_i^2-1)\sigma_z^{ii}
\end{eqnarray}
where $\sigma_z$ is the Pauli matrix.
 The action $S$ consists of two parts: the one describing the
 scattering of electron in the leads, and the other ($S_{HS})$ responsible
 for the fluctuating Habbard-Stratonovich fields
 \begin{eqnarray}\label{sof}
    S&=&-iTrLnG^{-1}+S_{HS}\\
G^{-1}(t,t')&=&[i\frac{\partial}{\partial
t}-\tilde{\epsilon}-\frac{\Gamma
b^2}{2\pi}\sigma_z(\tilde{g}_L+\tilde{g}_R)]\sigma_z\\
\tilde{g}_{L,R}(t,t')&=&e^{\pm
i\hat{\varphi}(t)}g_{L,R}(t-t')e^{\mp i\hat{\varphi}(t')}
,\,\,\label{glr} \\
 S_{HS}&=&\frac{C}{2e^2}\int dt (tr\dot{\hat{\varphi}}\sigma_z
 \dot{\hat{\varphi}})\\
 S_s &=& \frac{1}{e^2R_s }[\frac{i}{2}\int\int dt
 dt'\chi(t)\gamma(t-t')\chi(t')\nonumber\\
 &+&\int dt\chi(t)(eV-(\dot{\varphi}_1+\dot{\varphi}_2)/2)]
\end{eqnarray}
here $\Gamma$ is a tunnelling width, $\chi=\varphi_1-\varphi_2$,
the Fourier transform of  $\gamma(t)$  at finite temperature T is
 $\gamma(\omega)=\omega coth(\omega/2T)$. Here and below
the values with the hat $\hat{v},\hat{\varphi}$ mean the diagonal
$2\times2$ matrices with entries $v_1, v_2 $ and $ \varphi_1,
\varphi_2 $, respectively. $g_{L,R}$ represent Keldysh matrices of
the electron propagators in the leads at the position of the dot.
Their Fourier transforms are known: $g^r=-i\pi$,
 $g^{12}(\epsilon)=2i\pi f(\epsilon)$ with $f(\epsilon)$ as Fermi
distribution function. All dependencies on constant bias and
fluctuating potentials are collected in the exponents of
Eq.(\ref{glr}). We also define $T_{k}=\Gamma b^2$ as an effective
Kondo temperature. The $G(\omega)$ includes the Lagrange
multiplier which shifts the localized level position $\epsilon$ to
$\tilde{\epsilon}=\epsilon+\lambda $. For our purpose it is enough
to find both free parameters,
 $T_{k}$ and renormalized level $\tilde{\epsilon}$, by solving two
 self-consistent equations in equilibrium (neglecting the fluctuations of electrical
 potential) \cite{hew}. These equations follow from extremum
 conditions of effective action
  $S_{eff}$ relative to $b$ and $\tilde{\epsilon}$
\begin{eqnarray}\label{con}
  \frac{\partial S_{eff}}{\partial b} &=&
  \frac{\partial S_{eff}}{\partial\tilde{\epsilon}}=0
\end{eqnarray}
Unlike the multichannel Kondo model, the solution for  spin 1/2
Kondo quantum dot satisfies  \cite {hew} $\tilde{\epsilon}\sim T_K
(T_K /\Gamma)<<T_K$ which permits us to use $
\tilde{\epsilon}/T_K$ as a small parameter.

To obtain the effective action at one loop level we have to expand
the logarithm in equation (\ref{sof}) to the second power in
quantum field $\chi$. The calculations are straightforward,
though, rather long. The results for the unitary limit are in
accordance with those obtained by bosonization technique
\cite{matveev,aleiner2,brouwer,gogolin} and in reference
\cite{gol}. The part of effective action ($\sim Ln$) (\ref{sof})
which describes the electron scattering thus acquires a form
\begin{eqnarray}\label{s}
 \tilde{S} &=& \frac{N}{\pi }[\frac{i}{2}\int\int dt
 dt'\chi(t)\gamma(t-t')\chi(t')\nonumber\\
&-&\int dt\chi(t)(\dot{\varphi}_1+\dot{\varphi}_2)/2]
\end{eqnarray}
Here the number of the channels $N=2$. Collecting together all
terms we arrive to a compact form of effective action which is the
principal result of this section:
\begin{equation}\label{cg}
S_{eff} = \int\int dt
dt'\hat{\varphi}(t)\textbf{B}^{-1}(t-t')\hat{\varphi}(t')
\end{equation}
The Fourier image of Coulomb Green function $\textbf{B}(t)$ we
find $\textbf{B}(\omega)=\textbf{B}_v(\omega)/\omega^2$ where
\begin{eqnarray}\label{cb}
\textbf{B}_v(\omega)=\frac{E_c}{\omega^2+Q^2}\left(%
\begin{array}{cc}
  \omega^2-iQ\gamma & iQ(\omega-\gamma) \\
  -iQ(\omega+\gamma) & -\omega^2-iQ\gamma \\
\end{array}
\right)
\end{eqnarray}
here $Q=E_c \bar{N}/\pi$, $\bar{N}=N(1+h/(4e^2R_s))$ is the
effective number of conducting channels, and $\gamma\equiv
\gamma(\omega)$. $\textbf{B}_v(\omega)$ is matrix Green function
(GF) of electric potentials $v_i $. Having the effective action we
can calculate correlation functions of fluctuating potential
fields and, thus, find the conductance and shot noise corrections
in the Fermi liquid regime of the Kondo model.

{\it Fermi Liquid Regime}: Near the unitary limit it is convenient
to follow the scattering approach. For this purpose we use the
basis of s and p scattering states rather than those of the
left-lead and right-lead states. At $V=0$ the p states are
decoupled from the dot. For non-zero fluctuating potentials the
p-states are not decoupled, however, the s, p basis is important.
In real junctions with strong Kondo effect, nearly $T=0$  the dot
is described by strongly interacting fix-point with many body
state $(\bar{b})$. Before scattering the phases of both states,
$\bar{b}$ and s -state, coincide, while passing the scattering
region $\bar{b}$ state acquires only an extra phase $\pi$ compared
to that of s-state. Then the fix-point Hamiltonian can be written
\cite{glazman, pust, al} in the new basis $(b_{\sigma,k},
a_{\sigma,k})$ (here $a_{\sigma,k}$ stands for p-states) as $H=H_0
+H_s+H_{int}$
\begin{eqnarray}
H_0&=&\sum_{k,\sigma}\xi_{k}(b^{\dagger}_{\sigma,k}b_{\sigma,k}+
a^{\dagger}_{\sigma,k}a_{\sigma,k})\nonumber\\
&+&\frac{v}{2}\sum_{k,\sigma}(b^{\dagger}_{\sigma,k}a_{\sigma,k}+
a^{\dagger}_{\sigma,k}b_{\sigma,k})\nonumber\\
H_s&=&-\frac{a}{\nu
T_K}\sum_{k,k',\sigma}(\xi_{k}+\xi_{k'})b^{\dagger}_{\sigma,k}b_{\sigma,k'}\label{hs}\\
H_{int}&=&\frac{b}{\nu^2
T_K}b^{\dagger}_{\uparrow}b_{\uparrow}b^{\dagger}_{\downarrow}b_{\downarrow}\label{hi}
\end{eqnarray}
where $b=2a$ and $a=1/(2\pi)$ \cite{pust} and
$b_{\uparrow}=\sum_{k}b_{\uparrow,k}$. The Kondo temperature $T_K
$ is the only energy scale of the fixed-point Hamiltonian. We can
integrate out the $a_{\sigma,k}$-states and obtain the effective
action on Keldysh contour only for interacting states $\bar{b}$
\begin{eqnarray}\label{bs}
S&=&\int\int dtdt'\sum_{k\sigma}
b^{\dagger}_{k\sigma}(t)G_{bk}^{-1}(t-t')b_{k\sigma}(t')\nonumber\\
&-&\int dt(H^{i}_s +H^{i}_{int})\sigma_z^{ii}\\
G_{bk}^{-1}(t,t')&=&[(i\frac{\partial}{\partial t}-\xi_k)\sigma_z
-\frac{1 }{4}\hat{v}\sigma_z G_{ak}\sigma_z\hat{v}]\label{gin}
\end{eqnarray}
Here $G_{ak}$ is matrix Green function for decoupled
noninteracting $a$- states (for example,
$G^r_{ak}(\epsilon)=1/(\epsilon-\xi_k+i\delta)$). We separate
potentials in (\ref{gin}) into fluctuating part and constant
applied voltage $V$, that is, write $\hat{v}\rightarrow
V+\hat{v}$.
 To obtain the Coulomb interaction corrections to conductance and shot noise we
have to express the current in terms of Green functions (the
inverse of (\ref{gin}) and  then, with help of effective action
(\ref{cg}), to average the product of fluctuating potentials. This
goal can be achieved by applying  the perturbation theory (it
corresponds to 1/N expansion).

{\it Current}: To begin with, we consider, at first, the average
current. The
 backscattering current operator is given as \cite{glazman,gogolin1,me,oreg}:
\begin{eqnarray}
I_b&=&-\frac{ie}{\hbar}\frac{a}{2\nu
T_K}\sum_{k,k',\sigma}(\xi_{k}+\xi_{k'})
(b^{\dagger}_{\sigma,k'}a_{\sigma,k}-
a^{\dagger}_{\sigma,k}b_{\sigma,k'})\nonumber\\
&-&\frac{ie}{\hbar}\frac{b}{2\nu^2
T_K}\sum_{\sigma}(b^{\dagger}_{\sigma}a_{\sigma}-
a^{\dagger}_{\sigma}b_{\sigma})n_{\bar{\sigma}} \label{cur}
 \end{eqnarray}
 and $n_{\sigma}=b^{\dagger}_{\sigma}b_{\sigma}$,
 $\bar{\sigma}=-\sigma $. The sum over
 $\sigma $ stands for spin summation, which in our case is
 trivial.
 The averaged backscattering current
  consists of two parts
 $I_b=I_{s} +I_{int}$, where $I_{s}$ and $I_{int}$
 originate, respectively,
from the scattering  Hamiltonian $H_s$ and the interaction
$H_{int}$. $I_s$ and $I_{int}$ are generally expressed in terms of
Green's functions for $a$ and $\bar{b}$ states.
\begin{eqnarray}
I_{s}&=&-\frac{e}{\hbar}(\frac{a}{2\nu
T_K})^2\sum_{k,k',\sigma}(\xi_{k}+\xi_{k'})^2(A_{k,k'}-\bar{A}_{k,k'})\label{iss}\\
I_{int}&=-&\frac{e}{\hbar}(\frac{b}{2\nu
T_K})^2\sum_{k,\sigma}(\Pi_{k}-\bar{\Pi}_{k})
 \label{curi}
 \end{eqnarray}

\begin{figure}
\begin{center}
\includegraphics [width=0.45 \textwidth ]{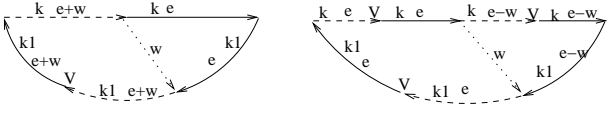}
\caption { Diagrams for processes which contribute to $I_s$. The
solid lines with momentum $k$ denotes $\bar{b}$-electron Green
functions $g_{bk}(\epsilon)$. Dashed lines represent GF of
noninteracting states $a$. The dot line stands for Coulomb GF
$iB_{v}(w)$ (\ref{cb}). All Green functions are $2\times2$ Keldysh
GF. Bias voltage $V$ denotes the vertex that connects dash and
solid line. Each vertex (but most left one) includes $\sigma_z$,
and  matrix product is considered.}
\end{center}
\end{figure}
here
\begin{eqnarray}
A_{k,k_1}&=&<[G_{ak}\hat{v}\sigma_zG_{bk}\sigma_zG_{bk_1}]^{11}_{tt}>\\
\bar{A}_{k,k_1}&=&<[G_{bk_1}\sigma_zG_{bk}\hat{v}\sigma_zG_{ak}]^{11}_{tt}>\nonumber\\
\Pi_{k}&=&<[G_{ak}\hat{v}\sigma_zG_{bk}\sigma_zG_{b}]^{11}_{tt}[G_b G_b]^{11}_{tt}>\\
\bar{\Pi}_{k}&=&<[G_{b}\sigma_zG_{bk}\hat{v}\sigma_zG_{ak}]^{11}_{tt}
[G_b G_b]^{11}_{tt}>\nonumber
 \end{eqnarray}
 where brackets $<...>$ denote the average over fluctuating
 potentials with the effective action (\ref{cg}). The expression
 in square brackets is matrix product in
 Keldysh space and convolution in time. One has to take the $11$
 component of this matrix product.
The last expression (\ref{curi}), unlike  the scattering part of
the current (\ref{iss}), includes, in addition, the Green
functions at the dot position $G_b=\sum_k G_{bk}$.  In the zero
order approximation (no potential fluctuations)
 for the Fermi liquid regime of Kondo effect
  the nonlinear
current was found before \cite{glazman, pust,al} (see also recent
works \cite{me,oreg,gogolin1}, where the explicit separation
between scattering and interacting contributions to conductance
was done). In this case the  backscattering current reads as,
\begin{equation}\label{con0}
     I_b ^0 =\frac{2e^2V}{3h}[ \pi^2(a^2+b^2 \frac{5}{4})(\frac{eV}{T_K
    })^2]
\end{equation}
 Let us now include
the fluctuations of potential fields. At first we do  the
calculations for $I_s$. For this purpose we expand $G_b$ to the
second order in fluctuating electrical potentials. The GFs for
$\bar{b}$-states to zero approximation  in fluctuating fields
acquire simple form:
\begin{figure}
\begin{center}
\includegraphics [width=0.4 \textwidth ]{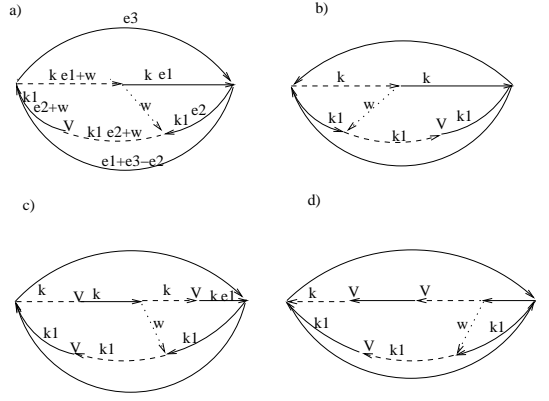}
\caption { The processes that contribute to the interacting part
of transport current $I_{int}$. The notations for momentum
dependent GF are the same as in Fig1. The lines without momentum
denote integrated over momentum GFs $g_b $.}
\end{center}
\end{figure}
\begin{eqnarray}
g_{bk} ^{R}(\omega)&=& \frac{\pi
i}{2}(\frac{1}{\omega-\xi_{k}-V/2+i\delta}+\frac{1}{\omega-\xi_{k}+V/2+i\delta}]\nonumber\\
g_{bk}^{12}(\omega)&=&-f(\xi_{k})(g ^{R}(k \omega)-g
^{A}(k\omega))\nonumber\\
g_{bk} ^{21}(\omega)&=&(1-f(\xi_{k}))(g ^{R}(k \omega)-g
^{A}(k\omega))\nonumber\\
g_b ^{i,j}(\omega)&=&\int d\xi_{k}g_{bk} ^{i,j}(\omega)\nonumber
 \end{eqnarray}
 The diagrams are a convenient way to
represent different contributions to the transport current. For $
I_s$ two types of diagram are shown in Fig2. The principal
contributions to the scattering part of the current at $1/N$ order
are given by a number of processes. These processes are presented
by: (i) two diagrams in Fig2, (ii) two diagrams similar to those
in Fig2, however, with the Coulomb (dot) line connected with other
vertex (V) (these diagrams are obtained by an interchange between
the V vertex and the end of the dot line ), and, (iii) four
"conjugate" diagrams for $\bar{A}_{k,k'}$ (like the cases  (i) and
(ii)). These "conjugate" diagrams
 together with diagrams in Fig2 simply double the real  parts
of those in Fig2. We notice that there are yet processes which can
be find by self-energy insertion in the one momentum line.
However, their impact on current is small. Thus summing all
relevant contributions and including the part without fluctuations
one finds the scattering part of the total current
\begin{equation}\label{js}
I_s=\frac{2e^2Va^2\pi^2}{3h}(\frac{eV}{T_K})^2[1
+\frac{9}{16\bar{N}}(\ln[\frac{2Q}{eV}]+1)]
\end{equation}
 We need more diagrams (see Fig3) to calculate $I_{int}$. Diagrams of types
 (a)and (c) are natural generalization of those in Fig2. Diagrams of type (b)
and (d) are of "scattering" type and appear only due to the
interactions. The accounting for relevant processes goes as
follow: at first the diagrams for $\bar{\Pi}$ are "conjugate to
those of $\Pi$ and after summation, as in the case for $I_s$, they
double the real parts of diagrams in Fig3. Next, we have :(i) four
diagram of type (a)(two of them appear  due to interchange with
vertex $V$), (ii) $2\bigotimes4=8$ diagram similar to (c)(factor
four appears due to four ways to connect V vertices),  (iii) two
diagrams of type (b), and (iv) four diagrams of type (d). The
summation all of this contributions results in the Coulomb
interaction correction to the $I_{int}$
\begin{equation}\label{jint}
I_{int}=
        \frac{2e^2Vb^2\pi^2}{3h}\frac{5}{4}(\frac{eV}{T_K})^2
        [1+\frac{2}{\bar{N}}(\ln[\frac{2Q}{eV}]+0.11)]
\end{equation}
The corresponding corrections to differential conductance due to
electron-electron interaction are obtained by taking the
derivative of (\ref{js}), (\ref{jint}) with respect to V.

{\it Shot Noise}: The effective charge of the current-carrying
particles is defined as $e^*=S_n /I_b$, where $S_n$ is the zero
frequency  Fourier transform of current-current correlation
function,
\begin{equation}
S(t)\equiv<I(t)I(0)>-<I>^2, \label{S}
\end{equation}
The same as the current, noise power can be separated  into the
two terms
 $S=S_n ^{s}+S_n ^{int}$ related, respectively, to the scattering
 part of the Hamiltonian $H_{s}$ and to the interaction part
 $H_{int}$. In the Fermi liquid regime (without fluctuating potentials)
at temperature $T=0$ the shot noise was obtained before
\cite{me,oreg}
\begin{equation}\label{nn}
    S_{n}^0=\frac{e^2V}{h}\pi^2(\frac{3}{2}b^2  +
    \frac{2a^2 }{3})\frac{(eV)^2}{T^{2}_K}
\end{equation}
\begin{figure}
\begin{center}
\includegraphics [width=0.4 \textwidth ]{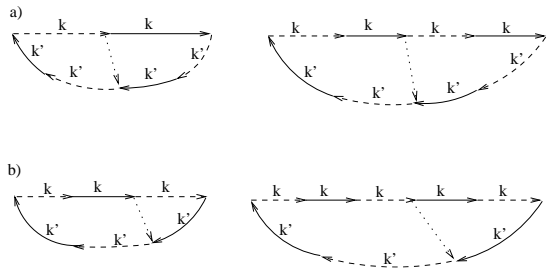}
\caption {  Diagrams which contribute into the scattering part of
the noise. Notations of the GFs and momentums are the same as in
Fig2, however, the time contour indices of the edge vertices are
different: here index $2$ is in the left and index $1$ in the
right vertex. Diagrams b) enter with sign minus.}
\end{center}
\end{figure}
The influence of Coulomb interaction changes the scattering and
interacting parts of $S_n$. Explicitly, we can find these changes
 (like in the case of the current) at $1/N$ order of perturbation theory.
 In the formula for noise (\ref{nn}) we assign Keldysh indices
 to the current operators (\ref{cur}), and then, using Keldysh technique, sum the all
  relevant diagrams.
  \begin{figure}
\begin{center}
\includegraphics [width=0.4 \textwidth ]{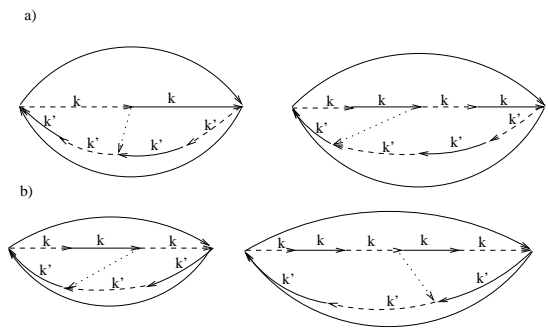}
\caption { The processes that contribute to the interacting part
of shot noise $S_n ^{int}$. The notations for GF correspond to
those in Fig2 and Fig3. The edge Keldysh vertex indices are the
same as in Fig3. Diagrams b) enter with minus sign.}
\end{center}
\end{figure}
   For scattering part
  $S_n ^{s}$ these diagrams are shown in Fig.4. There are
  also equivalent (not represented by Fig.4) diagrams. They also
  contribute  to the noise, and their number, like for current, can be
  easily counted. Thus, for the total scattering noise power
  which includes potential fluctuations, we obtain
\begin{equation}\label{ns}
S_{n}^s=\frac{2e^2Va^2\pi^2}{3h}(\frac{eV}{T_K})^2[1
+\frac{1}{2\bar{N}}(\ln[\frac{2Q}{eV}]+1.18)]
\end{equation}

For interacting part of the shot noise $S_n ^{int}$ the typical
diagrams are represented in Fig.5. In addition to them there are
also "scattering" diagrams (GFs of inside bobble point in one
direction). Each such diagram corresponds (with the same
contribution) to one in Fig.5. After summation of all diagrams we
obtain
\begin{equation}\label{nint}
S_{n}^{int}=
        \frac{3e^2Vb^2\pi^2}{2h}(\frac{eV}{T_K})^2
        [1+\frac{5}{6\bar{N}}(\ln[\frac{2Q}{eV}]+0.31)]
\end{equation}
By using Eqs.(\ref{ns},\ref{nint},\ref{js},\ref{jint}) and
definition of the $e^*$ we can immediately obtain factor $F$ (see
Eq.\ref{factor}).

 {\it Conclusions}: We have presented a theory to evaluate the
 effects of electron-electron interactions on quantum transport
 through the Kondo quantum dot. We address the important
 problem of effective backscattering charge
 for current-carrying particles in the interacting
 systems. This charge can be obtained from
 the measurements of shot noise and backscattering current in the Fermi liquid
 regime of a Kondo quantum dot. We show that in this regime
  the potential fluctuations
 can markable reduce the recently calculated universal value $5/3e$
 \cite{oreg}, though, the effective charge remains bigger then $e$.
 To get our principal result we have
 calculated the transport current and obtained the
 zero frequency shot noise power. Of cause, the potential fluctuations are not the only
 cause to change the effective charge. The pure unitary
limit in real system may by broken by the symmetry breaking
interactions like potential scattering or magnetic field which can
also influence the effective charge. However, in the case these
effects are small, they can be properly estimated
\cite{pust,oreg,al,me} and render a small correction to $e^*$.
\begin{acknowledgments}
I would like to thank A. D. Mirlin and A. Schiller
 for discussion I have had with them at the early stage of performing this work.
\end{acknowledgments}

\end{document}